  \documentclass[twocolumn,prb,aps]{revtex4}

 \usepackage[dvips]{graphicx}
 \usepackage[dvips]{graphics}

\newlength{\bxwidth}\bxwidth=0.8\textwidth

\begin{document}
\title{Bond-Bending and Bond-Stretching Phonons in Ferromagnetic La$_{0.7}$Sr$_{0.3}$MnO$_{3}$}
\author{D. Reznik$^{1,2\ast}$ and W. Reichardt$^1$}

\affiliation{$^1$Forschungszentrum Karlsruhe, Institut f\"ur Festk\"orperphysik, Postfach 3640, D-76021 Karlsruhe, Germany.\\
$^2$Laboratoire L\'eon Brillouin, CEA-CNRS, CE-Saclay, 91191 Gif sur Yvette, France.\\
}

\pacs{PACS numbers: 75.47.Lx, 63.20.Dj, 61.12.Ex }
% 63.20.Dj Phonon states and bands, normal modes, and phonon dispersion
% 75.47.Lx Manganites

\begin{abstract}

Longitudinal optical phonons with oxygen character were measured in La$_{0.7}$Sr$_{0.3}$MnO$_{3}$ by inelastic neutron scattering in the (1 0 0) cubic direction and results were compared with shell model predictions. Measurements were performed in several Brillouin zones, which enabled us to identify the eigenvectors independent of the shell model. All major disagreements between model predictions and experimental results are primarily due to the anomalous downward dispersion of the bond-stretching vibration. The main new result is that the rhombohedral distortion of the cubic lattice makes the bond-stretching vibrations interact strongly with bond-bending modes folded into the cubic Brillouin zone.

\end{abstract}

\maketitle

\section{Introduction}
Interplay between the charge, spin, and lattice degrees of freedom in perovskite manganites results in a multitude of unconventional properties of fundamental as well as practical interest. Ferromagnetic alignment of Mn core spins appears at some stoichiometries due to the double exchange interaction and the ferromagnetic-paramagnetic (FM) transition in these systems is accompanied by large magnetoresistance (MR). As predicted by theory\cite{Millis}, polarons forming and condensing above the FM transition temperature, T$_c$, have been observed by neutron scattering in La$_{0.7}$Ca$_{0.3}$MnO$_{3}$ and other manganites exhibiting colossal magnetoresistance (CMR)\cite{Adams,Dai,Vasiliu-Doloc}. Trapping of conduction electrons by these polarons is held responsible for the CMR effect. While the presence of polarons above T$_c$ demonstrates that electron-phonon coupling due to the Jahn-Teller effect is crucial to understanding the essential physics, relatively little is known about phonons in these materials.

"113" manganites such as La$_{1-x}$Ca$_{x}$MnO$_{3}$ and La$_{1-x}$Sr$_{x}$MnO$_{3}$ have a cubic perovskite structure except for a rotation of the MnO$_{6}$ octahedra, which doubles the unit cell.\cite{Dabrowski} In this paper we will use the notation based on the cubic unit cell of the undistorted structure, which makes the analysis easier to understand. Use of this notation requires folding in extra phonon branches that arise from the doubled unit cell. (For example, some phonons observed only at the zone boundary in the undistorted structure have nonzero structure factor at the zone center in the distorted structure, etc.) A previous study of oxygen phonons in La$_{0.7}$Sr$_{0.3}$MnO$_{3}$\cite{Reichardt} found anomalous downward dispersion of the bond-stretching vibration in the (1 0 0) direction, but it did not examine effects of interactions of these vibrations with the folded-in branches allowed by the tilt of the octahedra.

According to the presently accepted view of the physics of the ferromagnetic manganites, the strongest temperature dependence is expected of the Jahn-Teller modes, whose wavevectors are (0.5 0.5 0.5) and (0.5 0.5 0). But investigations of La$_{0.7}$Ca$_{0.3}$MnO$_{3}$\cite{Zhang} and La$_{0.7}$Sr$_{0.3}$MnO$_{3}$\cite{Reichardt2} found that these phonons did not have strong temperature dependence. Instead, a significant intensity loss on raising the temperature towards the FM transition was reported for phonons with wavevectors dispersing in the 1 0 0 direction. These results motivated us to go beyond the study of Reichardt and Braden.\cite{Reichardt} We measured phonon dispersions with higher resolution where possible and identified eigenvectors of \textbf{q}=(X 0 0) phonons in La$_{0.7}$Sr$_{0.3}$MnO$_{3}$ including interactions with the folded-in branches into the analysis. Although the crystals are twinned, the (1 0 0) direction is the same for the two domains.

\section{Experimental and Calculation Details}
The experiments were carried out on the triple-axis spectrometer 1T located at the ORPHEE reactor using doubly focusing Cu111 and Cu220 monochromator crystals and PG002 analyzer fixed at 13.7, 14.8, or 30.5 meV. Our samples were two high quality single crystals of La$_{0.7}$Sr$_{0.3}$MnO$_{3}$ with the FM transition temperature measured at 355K.  The volume of each crystal was ~0.5cm$^3$. The results were the same for the two crystals. All measurements were performed in the ferromagnetic phase at 100K (we did not go to lower temperatures for technical reasons). Since the FM transition temperature is much higher, we expect to get the same results at lower temperatures. This was confirmed by repeating some scans at 12K.

We performed measurements in several Brillouin zones and, guided by shell model predictions and symmetry constraints, found by trial and error approximate phonon eigenvectors that correctly give the observed intensities. In this procedure we ignored small transverse atomic displacements, since these would not affect longitudinal phonon intensities in the Brillouin zones that we measured. They should exist due to the rhombohedral distortion.

\section{Calculation Results}

Ignoring the rhombohedral distortion of the lattice, we expect to
observe two optical oxygen branches intrinsic to the cubic
perovskite structure above 35 meV. These are the bond-stretching
and bond-bending modes. The rhombohedral distortion doubles the
unit cell along one of the (111) directions and thus the number of
phonon branches is expected to double as well. Shell model
calculations show that the bond-bending and the bond-stretching
branches should survive in the distorted structure, plus two
additional branches of bond-bending character should be folded
into the cubic Brillouin zone due to the unit cell doubling. At
the zone center one of the folded-in modes has a vanishing
structure factor, whereas the other one has a vanishing structure
factor at the zone boundary. Thus there should be three modes with
nonzero intensity at the zone center and the zone boundary and
four modes in the middle of the zone. According to the model, the
original bond-bending branch should disperse upwards, whereas the
bond-stretching and the folded-in branches should be approximately
flat. (See Fig. 1)

Structure factors of the bond-bending and folded-in modes are very
sensitive to the value of Mn-O-Mn bond angle whereas the structure
factor of the bond-stretching vibration is not. The bond angle
depends on the magnitude of the rhombohedral distortion of the
lattice and is reported to be 164$^{\circ}$ for
La$_{0.7}$Sr$_{0.3}$MnO$_{3}$.\cite{Mitchell} In the absence of
the rhombohedral distortion, i.e. when the Mn-O-Mn bond angle is
180$^{\circ}$, all longitudinal phonon intensities should be
proportional to \textbf{Q}$^{2}$.

\begin{figure}[ptb]
%\centerline{\includegraphics[width= cm]{fig1.eps}} \caption{
\includegraphics[width=7 cm]{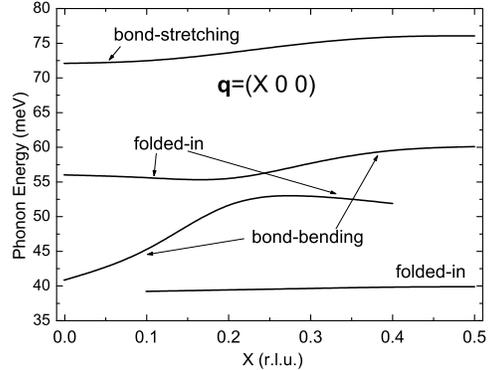}
\caption{
{\label{fig1}} Longitudinal oxygen phonon dispersions along the 1 0 0 cubic direction calculated using the shell model. Only the branches with nonzero structure factors are shown. }
\end{figure}

\section{Experimental Results}

\subsection{Zone Center}

\begin{figure}[ptb]
%\centerline{\includegraphics[width= cm]{fig2.eps}} \caption{
\includegraphics[width=7 cm]{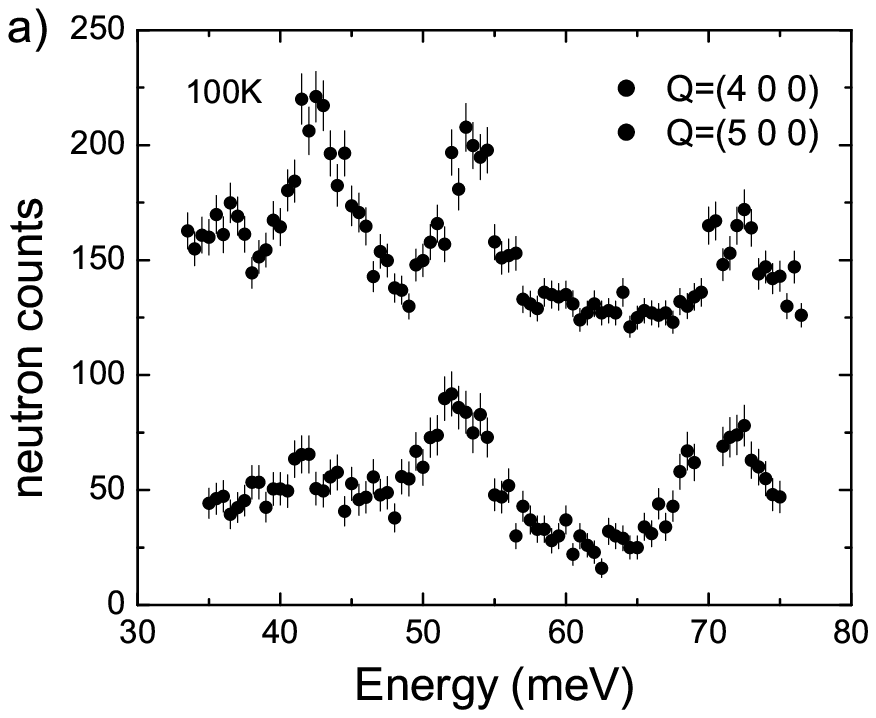}
\includegraphics[width=2 cm]{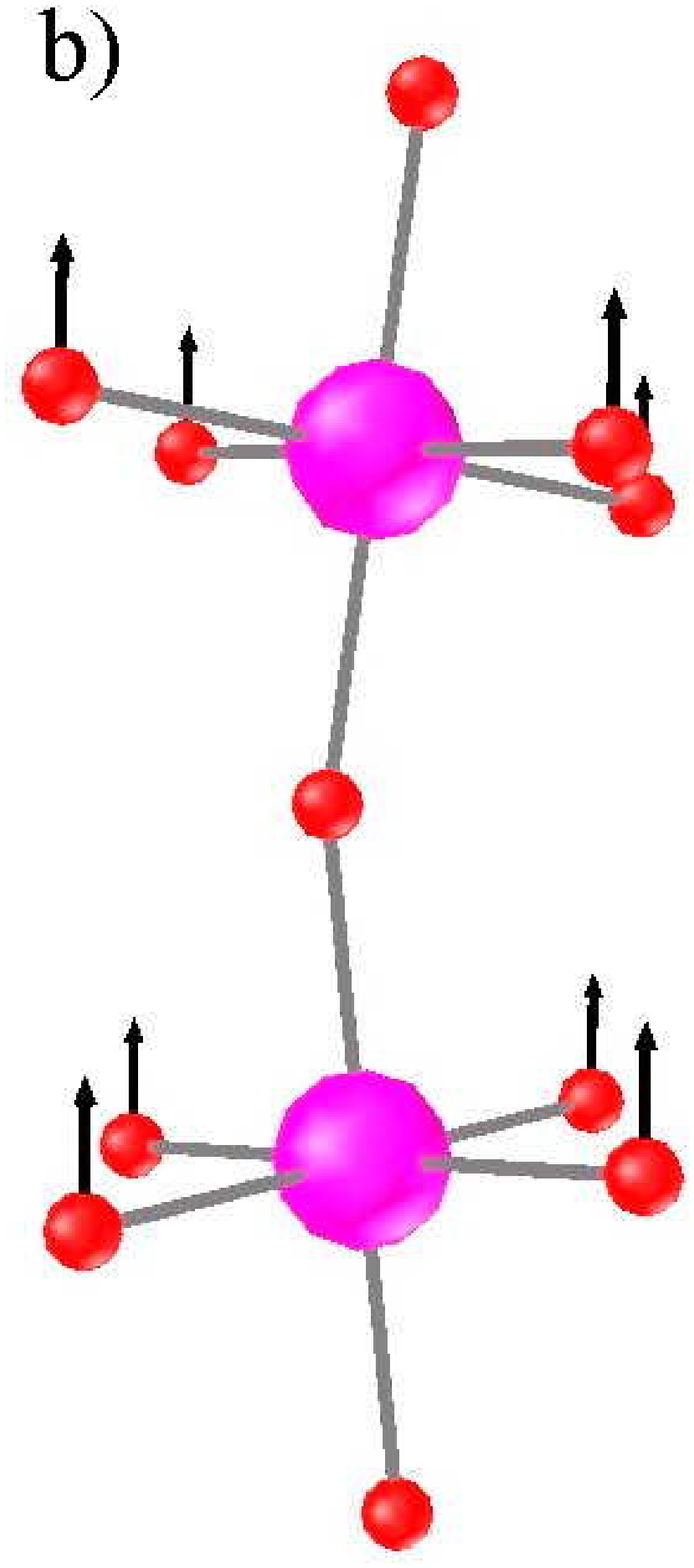} \includegraphics[width=2.1 cm]{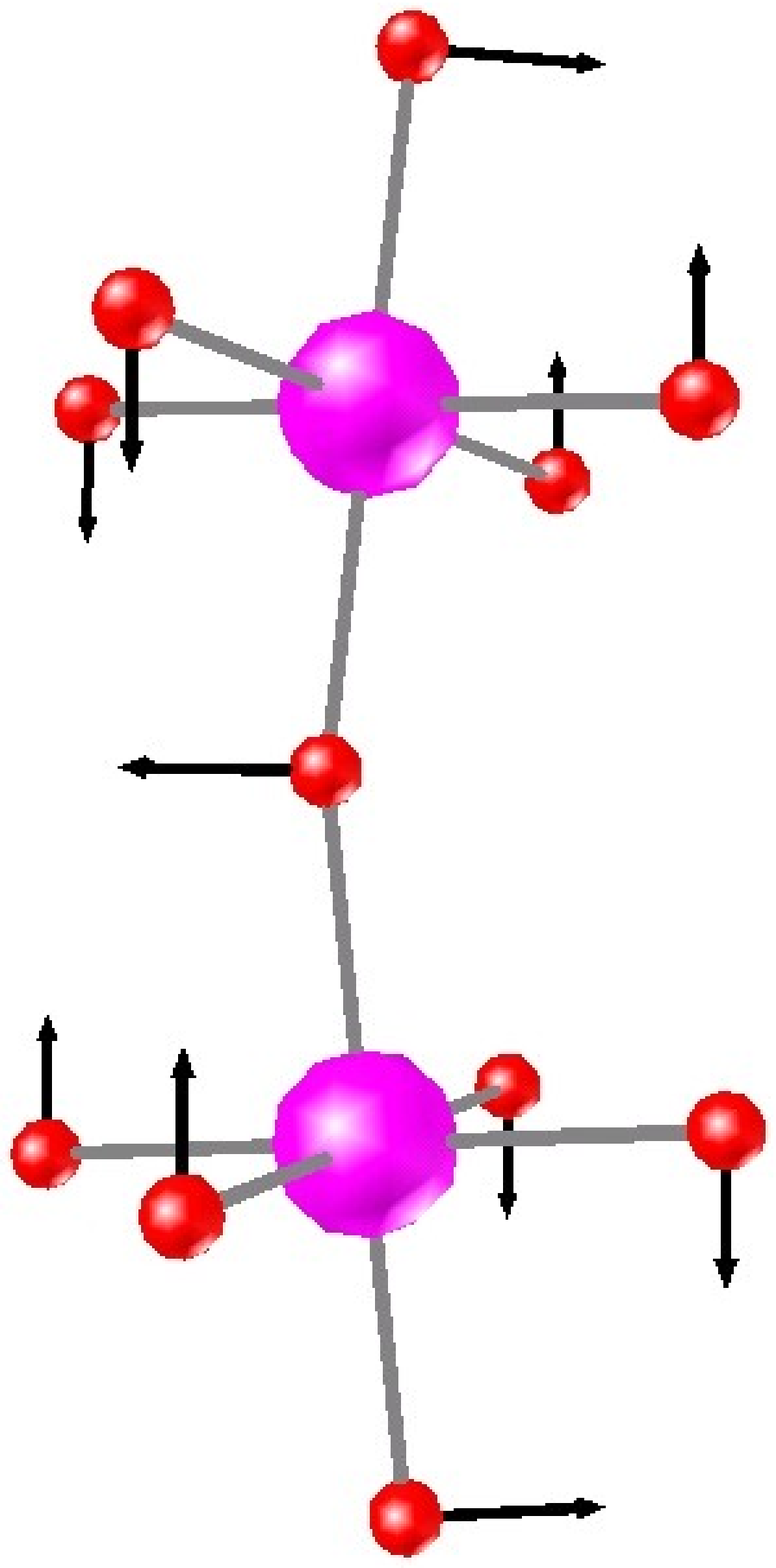} \includegraphics[width=2 cm]{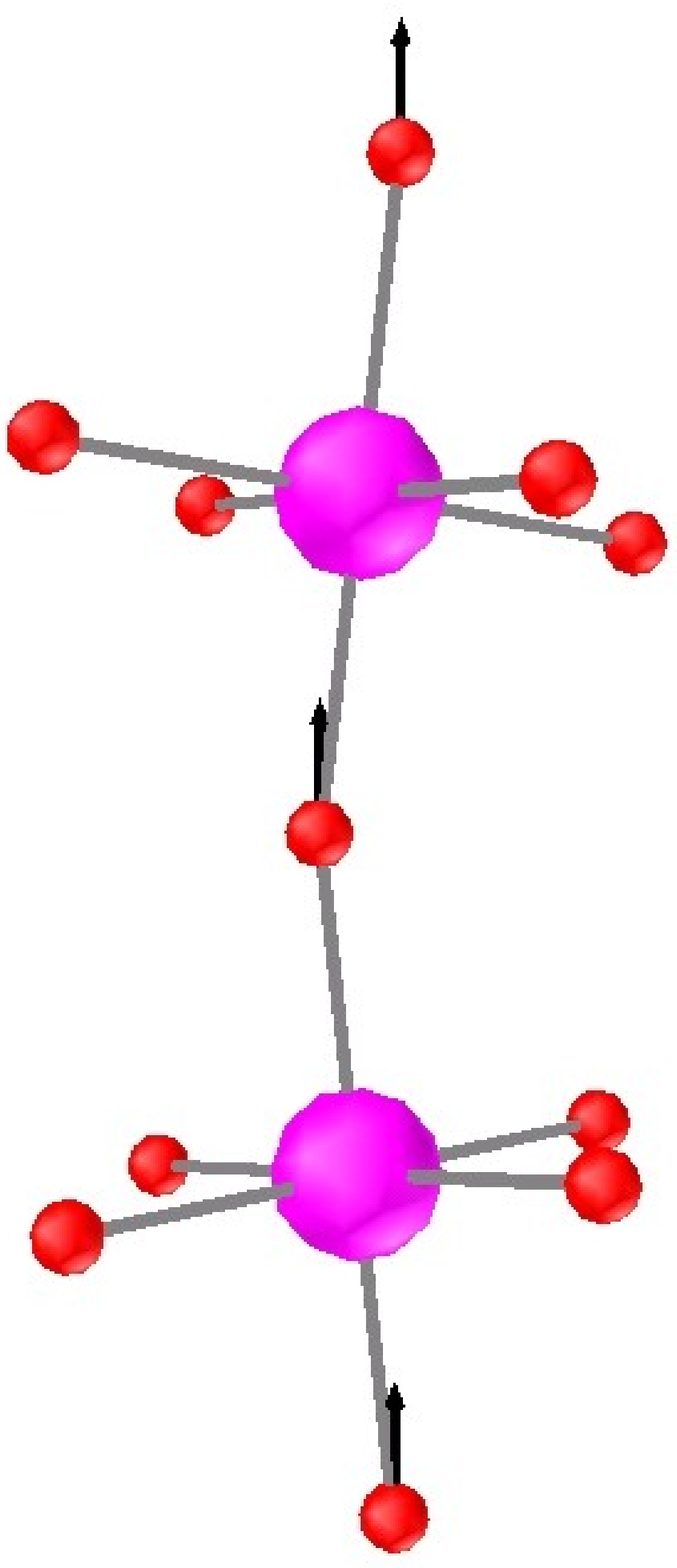}
\caption{ {\label{fig2}} (Color online) a) Zone center phonons
measured in two Brillouin zones. Resolution was lower at
\textbf{Q}=(5 0 0). b) Approximate eigenvectors of the zone
boundary modes at 100K deduced from the measured intensities. Mode
frequencies left to right are: 42mev, 53meV, 73meV. Only pairs of
MnO$_6$ octahedra are shown.}
\end{figure}

We observe the zone center phonons at 42meV, 53meV, and 72meV.
(Fig. 2a) According to the shell model these should be the
energies of the bond-bending, folded-in, and the bond-stretching
modes respectively (Fig. 2b). This assignment can be verified by
structure factor calculations for the corresponding polarization
patterns. The 42 meV bond-bending vibration decreases in intensity
by a factor of 1.82 (Fig. 2a) while the structure factor of the
corresponding (leftmost) polarization pattern in figure 2b gives
2.17. The increase of the 53meV phonon intensity from
\textbf{Q}=(4 0 0) to \textbf{Q}=(5 0 0) is 1.42 vs the calculated
value of 1.96. The 72meV mode intensity increases by a factor of
1.28 vs. the calculated 1.53. Considering that we ignore small
deviations from the cubic high symmetry directions as discussed
above as well as small admixtures of other modes, structure factor
calculations confirm our assignment.

\subsection{Zone Boundary}

\textbf{q}=(0.5 0 0) (zone boundary) is another high symmetry point of the reciprocal lattice. We performed measurements in different Brillouin zones in order to gain insight into the zone boundary phonon eigenvectors. Three longitudinal oxygen modes appear at 38, 47, and 61 meV.

\begin{figure}[ptb]
%\centerline{\includegraphics[width= cm]{fig2.eps}} \caption{
\includegraphics[width=7 cm]{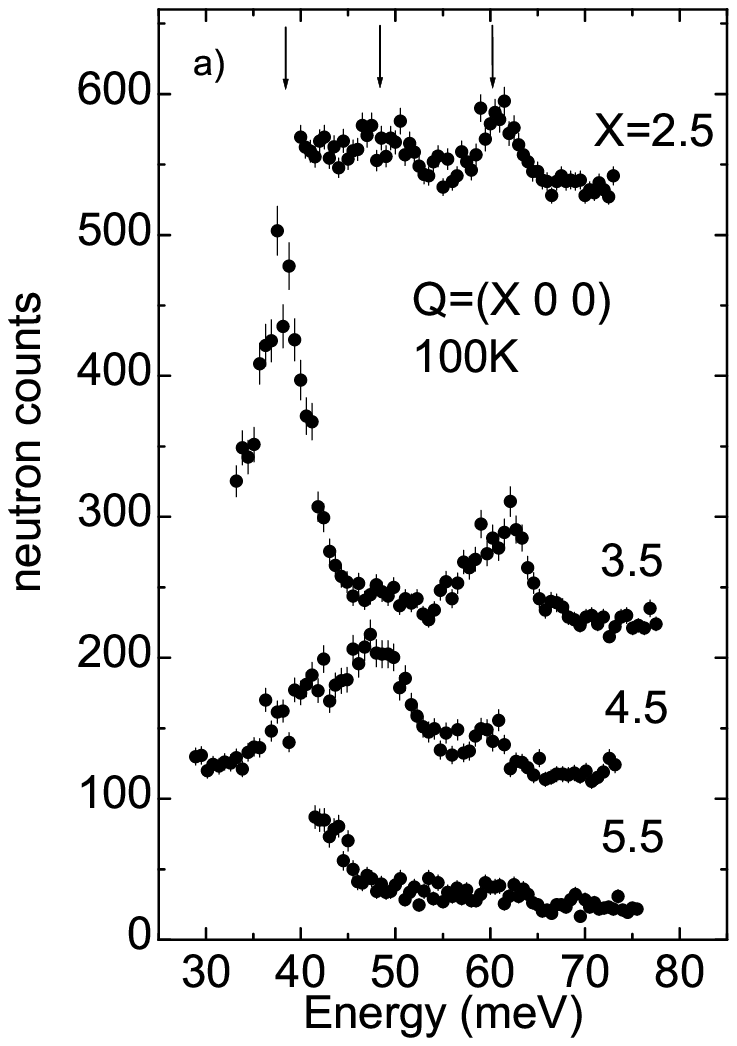}
\includegraphics[width=2 cm]{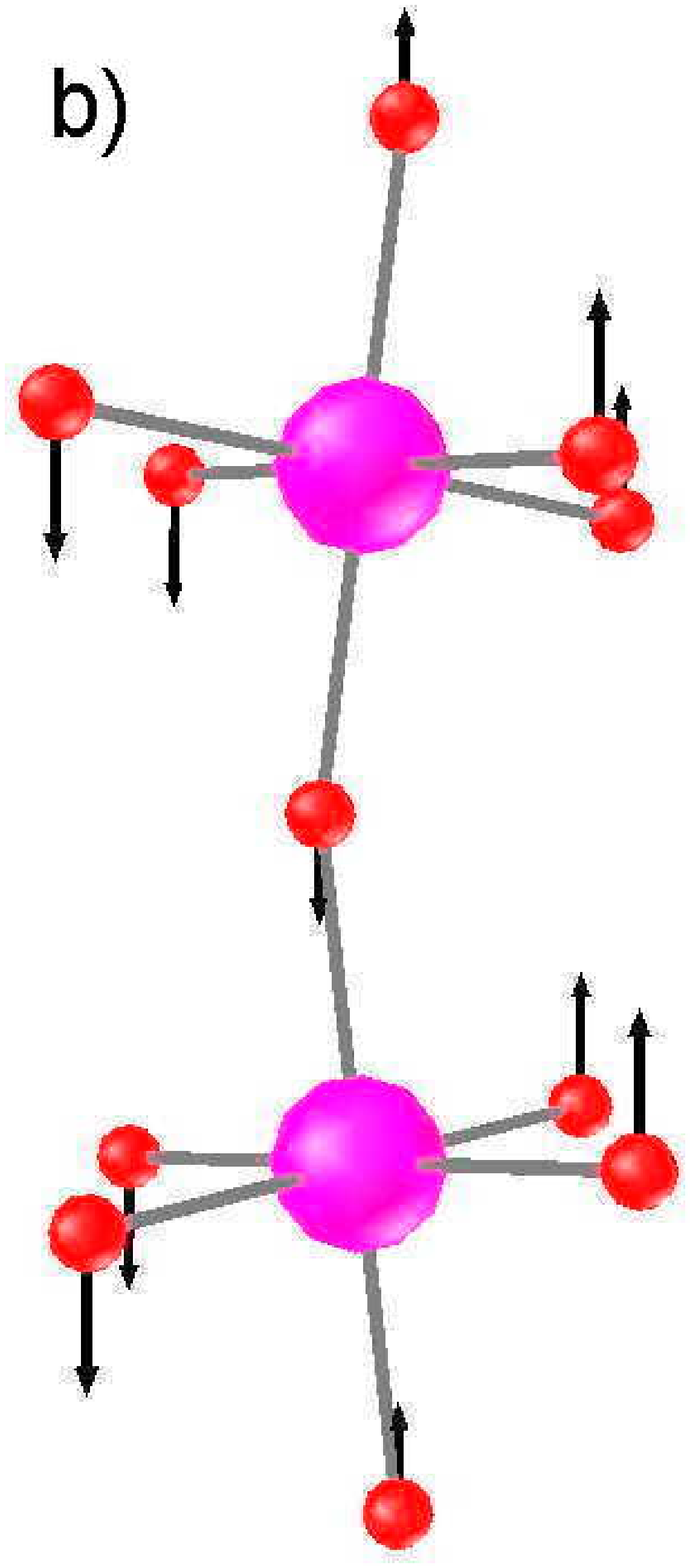} \includegraphics[width=2.1 cm]{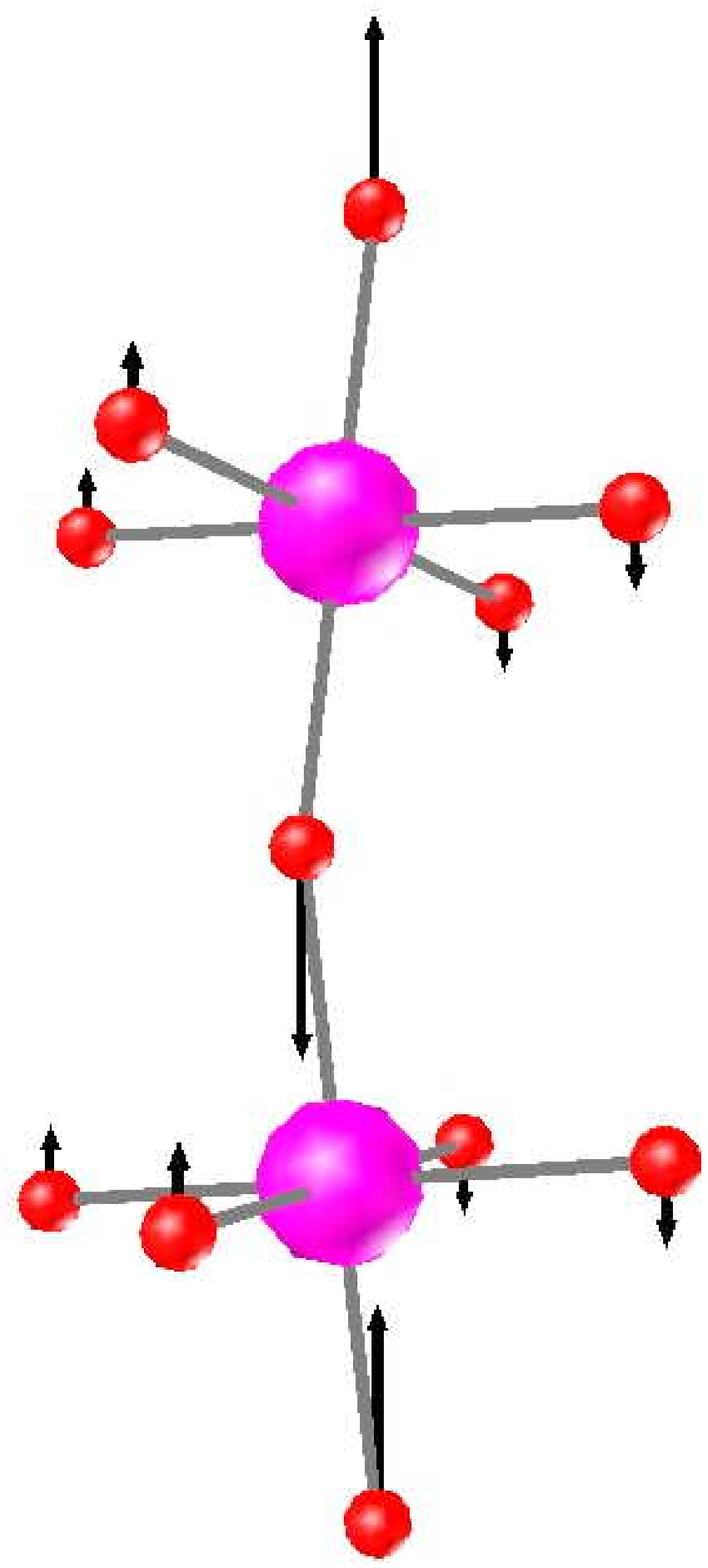} \includegraphics[width=2 cm]{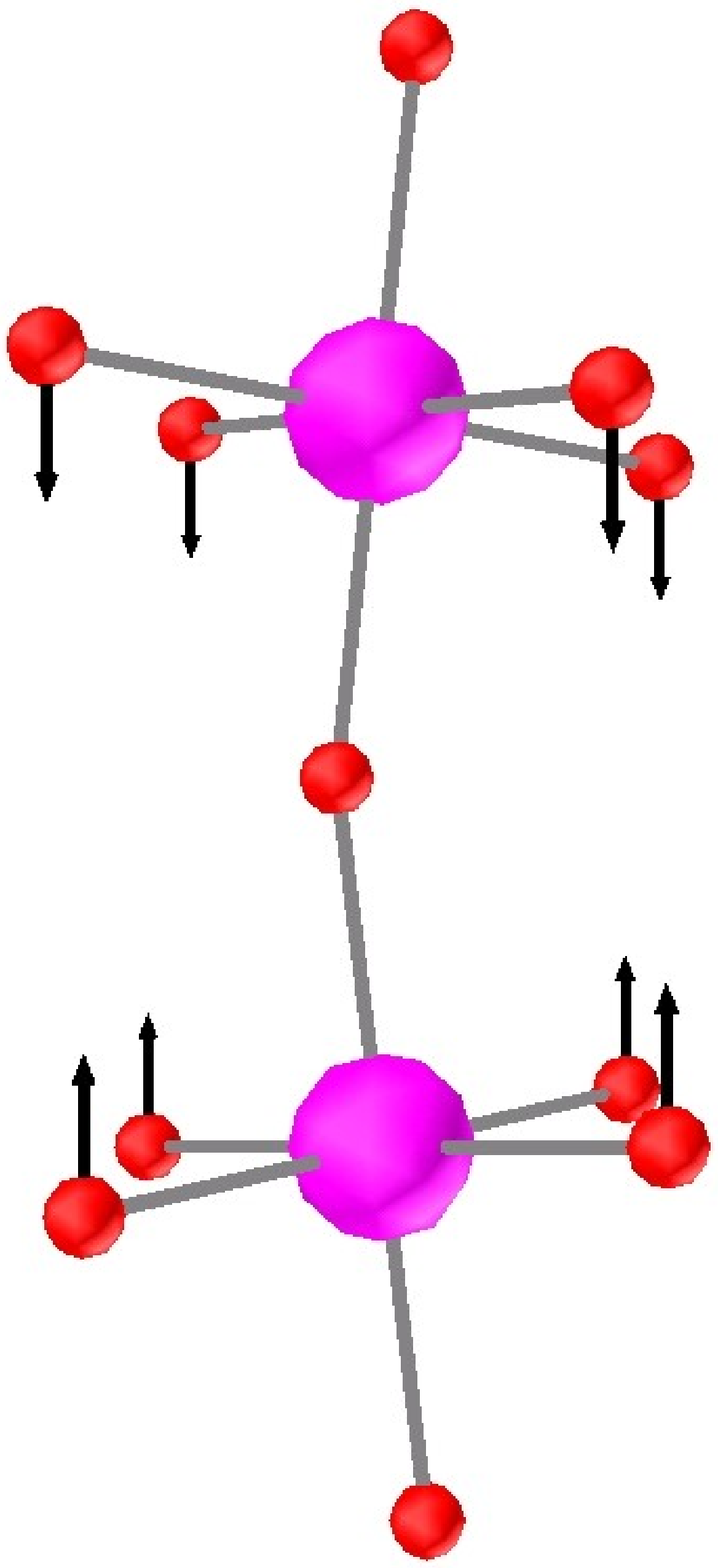} \caption{
{\label{fig3}} (Color online) a) Zone boundary phonons measured in
four Brillouin zones. b) Approximate eigenvectors of the zone
boundary modes at 100K in the MnO$_6$ octahedra deduced from the
measured intensities. Mode frequencies left to right are: 38mev,
47meV, 61meV.}
\end{figure}
\begin{figure}[ptb]
%\centerline{\includegraphics[width= cm]{fig2.eps}} \caption{
\includegraphics[width=8 cm]{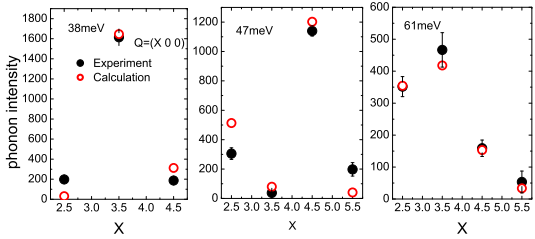}
\caption{ {\label{fig4}} (Color online) Comparison of the measured
and calculated phonon intensities at the zone boundary for the
38meV, 47meV, and 61meV phonons. The calculated values are the
predicted intensities assuming the corresponding polarization
patterns in Fig. 3b. An arbitrary scaling factor (the same for all
points in the three graphs) was applied to the calculated values.}
\end{figure}

We begin the analysis with the 61 meV phonon. Its intensity decreases with increasing \textbf{Q} and almost vanishes at \textbf{Q}=(5.5 0 0). That is precisely what is expected from the bond-bending zone boundary mode (rightmost polarization pattern in figure 3b) as discussed for the zone center modes. In fact structure factor calculations for this phonon eigenvector are in good quantitative agreement with the observed intensity at 61meV (Fig. 4). The shell model predicts the correct frequency for this mode, thus we can assign it with a very large degree of confidence.

The 47 meV phonon is the strongest at \textbf{Q}=(4.5 0 0); it is weaker at \textbf{Q}=(2.5 0 0) and is absent at \textbf{Q}=(3.5 0 0) and \textbf{Q}=(5.5 0 0) (Fig. 3a). Structure factor calculations based on the middle polarization pattern in figure 3b correctly reproduce the observed intensity at all measured \textbf{Q}. (Fig. 4)  The 47 meV phonon is mostly bond-stretching in character, but with a significant bond-bending component of folded-in character (The bond-bending component in the 47 meV mode is orthogonal to the bond-bending vibration at 61 meV).

The 38 meV phonon appears most strongly at \textbf{Q}=(3.5 0 0), it is weaker at \textbf{Q}=(5.5 0 0) and \textbf{Q}=(4.5 0 0), and is very weak at \textbf{Q}=(2.5 0 0) (Fig. 3a). The leftmost polarization pattern in figure 3b gives the correct intensities at all measured \textbf{Q}. (Fig. 4) Just like for the 47 meV phonon, its eigenvector is a superposition of the bond-stretching and folded-in bond-bending vibrations, but here the bond-bending vibrations have a larger amplitude.

The coupling between the bond-bending component and the bond-stretching one in the 38 and 47meV modes is exclusively due to the tilt of the octahedra and should decrease with increasing Mn-O-Mn bond angle. Most of the bond-stretching character at the zone boundary is at 47 meV with a smaller contribution at 38 meV. These energies are much lower than the prediction of the shell model and indicate strong electron-phonon coupling.

\subsection{Between the Zone Center and the Zone Boundary}

Deducing the eigenvectors between the zone center and the zone
boundary as was done at the zone center and the zone boundary is
much more difficult due to lower symmetry. However, on a
qualitative level, this analysis is still useful. Figure 5 shows
the phonon spectra at three reduced reciprocal lattice vectors
each measured in two Brillouin zones. There are four phonon peaks
in agreement with the predictions of the shell model.

We start with the bond-bending branch, which is the easiest one to
identify by its intensity decrease towards large \textbf{Q} (see
above). The intensity of the highest energy mode at X=0.4 and 0.3
decreases from \textbf{Q}=(4-X 0 0) to \textbf{Q}=(4+X 0 0). (Fig.
5) It is also very strong at \textbf{Q}=(2.3 0 0) (data not
shown). Such behavior is similar to the 61 meV mode at the zone
boundary, thus we conclude that the highest energy phonons at
X=0.4 and 0.3 are of bond-bending character. The bond-bending
branch originates at 42 meV at the zone center (Fig. 2) and
disperses upwards crossing the 53 meV branch at
\textbf{q}$\approx$(0.15 0 0)(data not shown). Mixing with the
bond-bending branch may be the reason that the 53meV peak is
stronger at \textbf{Q}=(3.8 0 0) than at \textbf{Q}=(4.2 0 0).
Between \textbf{q}=0.1 and \textbf{q}=0.3 it hybridizes with the
other branches emerging again as a pure mode at \textbf{q}=(0.3 0
0) - (0.5 0 0) around 60 meV.

\begin{figure}[ptb]
%\centerline{\includegraphics[width= cm]{fig2.eps}} \caption{
\includegraphics[width=7 cm]{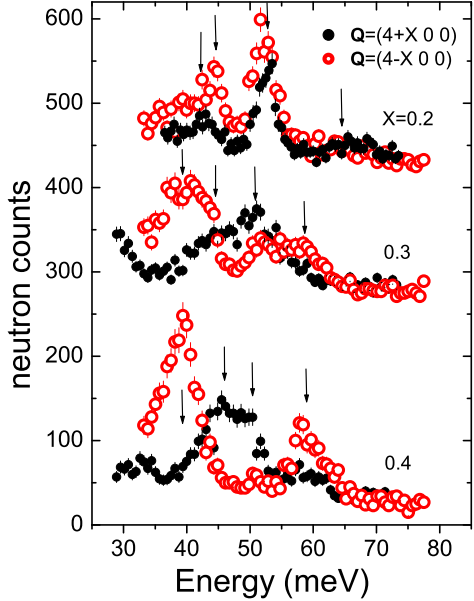}
\caption{ {\label{fig5}} (Color online) Phonons between the zone
center and the zone boundary measured in two Brillouin zones}
\end{figure}

Identification of the dispersion of the bond-bending branch allows us to assign the other modes to either the bond-stretching or the folded-in branches. The mode at \textbf{q}=(0.2 0 0) and 65 meV has mostly bond-stretching character, since its intensity is stronger at \textbf{Q}=(4.2 0 0) than at \textbf{Q}=(3.8 0 0). We also know that the mode at \textbf{q}=(0.3 0 0) and 57meV is bond-bending. This indicates that the bond-stretching branch has a very steep downward dispersion between \textbf{q}=(0.2 0 0) and (0.3 0 0). This steep dispersion is responsible for the very large energy width of this mode at \textbf{q}=(0.2 0 0) due to the finite \textbf{q}-resolution of the spectrometer. At \textbf{q}=(0.3-0.5 0 0) the bond-stretching mode continues to disperse to lower energies, but not as steeply. In this range of wavevectors it hybridizes with the two folded-in branches. As a result we observe three peaks below 55 meV. All are of mixed folded-in/bond-stretching character as is clear from strong intensity changes between \textbf{Q}=(4-X 0 0) and \textbf{Q}=(4+X 0 0) (see discussion above for the zone boundary). Figure 6 illustrates the dispersion of the bond-stretching vibration (marked with arrows) in the data covering \textbf{Q}=(4-4.5 0 0). It disperses downward away from the zone center and at \textbf{q}$\geq$(0.2 0 0) it contributes to more than one mode due to its interaction with other branches.

\begin{figure}[tb]
%\centerline{\includegraphics[width= cm]{fig2.eps}} \caption{
\includegraphics[width=8.5 cm]{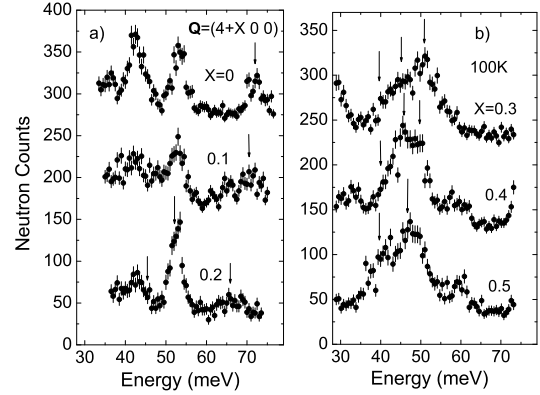}
\caption{ {\label{fig6}}  Evolution of the phonon spectra between
Q=4 0 0 and Q=4.5 0 0. Arrows mark modes with bond-stretching
character.}
\end{figure}
\section{Summary}

\begin{figure}[ptb]
%\centerline{\includegraphics[width= cm]{fig2.eps}} \caption{
\includegraphics[width=7 cm]{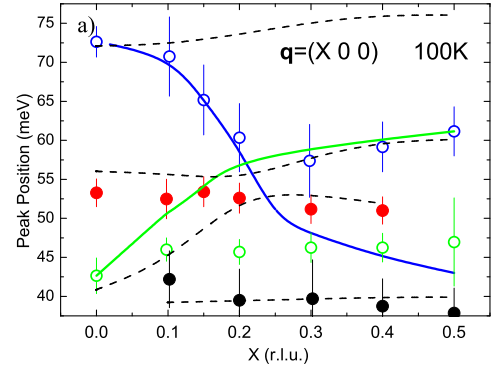}
\caption{ {\label{fig7}} (Color online) Circles represent measured
longitudinal phonon peak positions at 100K. Vertical bars
represent peak widths. Solid lines are guides to the eye showing
dispersions of the bond-stretching (downward-dispersing) and
bond-bending (upward-dispersing) vibrations in the absence of
interactions with other branches and each other. Their crossing of
each other and of the folded-in branches (see text) results in
complex phonon polarization patterns. Dashed lines show results of
the shell model calculations.}
\end{figure}

Figure 7 shows the dispersion of the longitudinal oxygen phonons
in the 1 0 0 direction summarizing the above results. The shell
model gives a good prediction for the frequencies of the
bond-bending and folded-in branches, however, it does not predict
a strong downward dispersion of the bond-stretching branch away
from the zone center as previously reported. \cite{Reichardt} This
anomaly has been observed in almost all metallic perovskite oxides
and is generally interpreted as a signature of strong
electron-phonon coupling. As a result of the downward dispersion,
the bond-stretching branch crosses the bond-bending branch at
\textbf{q}=0.2 and hybridizes with the folded-in branches at
\textbf{q}=(0.2-0.5 0 0). The bond-bending branch has an upward
dispersion in good agreement with the shell model. The folded-in
branches are nearly flat, but they are affected by the interaction
with the bond-stretching branch at \textbf{q}=(0.2-0.5 0 0).

\section{Conclusions}

In conclusion, we have demonstrated that the rhombohedral distortion of the lattice has a profound effect on the longitudinal oxygen phonons dispersing in the 1 0 0 cubic direction. It results in two nearly flat folded-in branches which couple strongly to the upward-dispersing bond-bending and the downward-dispersing bond-stretching branches. The latter appears as a pure mode only near the zone center hybridizing with the folded-in branches throughout most of the zone including the zone boundary.

\section{Acknowledgements}

The authors would like to thank L. Pintschovius for many helpful discussions and a critical reading of the manuscript.

\normalsize{$^\ast$To whom correspondence should be addressed;
E-mail: reznik@llb.saclay.cea.fr }


\begin{thebibliography}{10}

\bibitem{Millis}  A. J. Millis, Boris I. Shraiman, and R. Mueller, {Phys Rev. Lett.} {\bf 77},  175 (1996).

\bibitem{Adams} C. P. Adams, J. W. Lynn, Y. M. Mukovskii, A. A. Arsenov, and D. A. Shulyatev,  {Phys Rev. Lett.} {\bf 85}, 3954 (2000).

\bibitem{Dai} Pengcheng Dai, J. A. Fernandez-Baca, N. Wakabayashi, E. W. Plummer, Y. Tomioka, and Y. Tokura, {Phys Rev. Lett.} {\bf 85}, 2553 (2000).

\bibitem{Vasiliu-Doloc} L. Vasiliu-Doloc, S. Rosenkranz, R. Osborn, S. K. Sinha, J. W. Lynn, J. Mesot, O. H. Seeck, G. Preosti, A. J. Fedro, and J. F. Mitchell, {Phys Rev. Lett.} {\bf 83}, 4393 (1999).

\bibitem{Dabrowski}  B. Dabrowski, X. Xiong, Z. Bukowski, R. Dybzinski, P. W. Klamut, J. E. Siewenie, O. Chmaissem, J. Shaffer, C. W. Kimball, J. D. Jorgensen and S. Short , { Phys. Rev. B} {\bf 60}, 7006 (1999).

\bibitem{Reichardt}  W. Reichardt and M. Braden, { Physica B} {\bf 263B}, 416 (1999).

\bibitem{Zhang}  Jiandi Zhang, Pengcheng Dai, J. A. Fernandez-Baca, E. W. Plummer, Y. Tomioka, and Y. Tokura, { Phys Rev. Lett.} {\bf 86}, 3823 (2001).

\bibitem{Reichardt2}  W. Reichardt and M. Braden, unpublished.

\bibitem{Mitchell} J. F. Mitchell, D. N. Argyriou, C. D. Potter, D. G. Hinks, J. D. Jorgensen, and S. D. Bader, { Phys. Rev. B} {\bf 54}, 6172 (1996).

\end{thebibliography}
\end{document}